# Cervical Auscultation Machine Learning for Dysphagia Assessment

An An Chia*
Speech Therapy Department
Sengkang General Hospital
Singapore
chia.an.an@skh.com.sg

Stacy Lum*
Speech Therapy Department
Sengkang General Hospital
Singapore
stacy.lum@skh.com.sg

Michelle Boo*
Science Mathematics & Technology
Singapore University of Technology &
Design, Singapore
michelle_boo@mymail.sutd.edu.sg

Rex Tan
Aevice Health
Singapore
rex@aevice.com

Balamurali B.T
Science Mathematics & Technology
Singapore University of Technology &
Design, Singapore
balamurali_bt@sutd.edu.sg

Jer-Ming Chen
Science Mathematics & Technology
Singapore University of Technology &
Design, Singapore
jerming_chen@sutd.edu.sg

*These authors contributed **equally** to this work and are considered **co-first authors.**

*Abstract* — **This study evaluates the use of machine learning, specifically the Random Forest Classifier, to differentiate normal and pathological swallowing sounds. Employing a commercially available wearable stethoscope, we recorded swallows from both healthy adults and patients with dysphagia. The analysis revealed statistically significant differences in acoustic features, such as spectral crest, and zero-crossing rate between normal and pathological swallows, while no discriminating differences were demonstrated between different fluid and diet consistencies. The system demonstrated fair sensitivity (mean ± SD: 74% ± 8%) and specificity (89% ± 6%) for dysphagic swallows. The model attained an overall accuracy of 83% ± 3%, and F1 score of 78% ± 5%. These results demonstrate that machine learning can be a valuable tool in non-invasive dysphagia assessment, although challenges such as sampling rate limitations and variability in sensitivity and specificity in discriminating between normal and pathological sounds are noted. The study underscores the need for further research to optimize these techniques for clinical use.**

*Keywords* — ***Dysphagia, Swallowing Sounds, Machine Learning, Random Forest Classifier, Non-Invasive Diagnosis, Audio Signal Processing, Acoustic Feature, Cervical Auscultation***

## I. Introduction

Dysphagia refers to difficulties in swallowing, a symptom commonly associated with conditions such as stroke or head and neck cancer. Dysphagia can result in penetration and/or aspiration, which is when fluids or food enters the airway during swallowing. Consequences of dysphagia include aspiration pneumonia, malnutrition, dehydration, choking, and in extreme cases, death [1]. It is therefore important that patients with swallowing difficulties should be referred to a Speech Therapist (ST) for timely assessment and management.

To assess for dysphagia, ST's will typically perform a clinical bedside assessment. This often includes a case history interview, oromotor examination, and a series of swallowing trials using fluids and food. However, there may be variability in terms of sensitivity and specificity, which can lead to underdiagnosis or unnecessary interventions [2-3].

Videofluoroscopic Swallowing Study (VFSS), considered the "gold standard" assessment for diagnosing dysphagia [4], is a radiographic technique that visualizes the swallowing process by tracking a fluid or food bolus coated with a contrast agent, making it possible to reliably detect aspiration and the physiological function of the swallowing mechanism. However, VFSS involves radiation, specialized skills and equipment. This highlights a gap in current assessment methods underscoring the demand for a simple, reliable, and non-invasive method for evaluating swallowing function.

Cervical Auscultation (CA) is currently an adjunctive tool used during bedside swallowing assessments [5]. Utilizing a stethoscope, CA aims to identify dysphagia by listening to the respiration and sounds produced in the cervical area during swallowing. Despite its advantages of being portable, non-invasive and affordable, CA has variable accuracy, as reflected in the wide range of sensitivity (23% - 94%) and specificity (50% - 74%) [6]. This variability is often linked to the complex tissue and articulatory acoustics associated with these airway sounds, making it challenging to establish agreement among STs [7]. In addition, a normal human ear may not be sensitive enough to pick up on all the different breath-swallow sounds that could be produced as compared to a medical-grade sensor [8].

Given these constraints, recent studies have investigated the use of wearable sensors, advanced digital signal processing, and machine learning techniques, in detecting dysphagia. Two main categories of algorithms have been used: traditional machine learning and deep learning, with each approach offering distinct advantages and limitations.

Traditional machine learning algorithms, such as Support Vector Machines (SVM), Random Forests, and Hidden Markov Models, have been extensively employed in swallowing sound signal classification. Studies have demonstrated that the accuracy of traditional machine learning algorithms in binary classification tasks (76% to 99%) can exceed the accuracy of trained clinicians [9]. However, performance of these algorithms heavily depends on the type and quality of the selected signal and the extracted features. The latter was highlighted in two separate studies by Donohue et al [10,11] in which the differences in accuracy of detecting pathological swallow sounds across the two studies can be attributed to the different signal features selected.

On the other hand, deep learning algorithms, particularly Artificial Neural Networks (ANN) and Convolutional Neural

Networks (CNN), offer a more automated end-to-end approach, capable of extracting and classifying more abstract and advanced features without manual intervention. Deep learning methods have shown promise in swallowing sound classifications, but require large datasets for network training [9]. The effectiveness of deep learning algorithms becomes limited when the number of training sets is insufficient.

This study examines the potential of machine learning in identifying dysphagia through acoustic signals captured using a commercially available wearable stethoscope. We present the methodology used (participants, data collection protocol, signal processing and analysis), results obtained (statistical analysis of acoustic features; feature-space visualization; machine learning classification; and feature-importance analysis) which allows a discussion (on findings, machine learning performance and efficacy, and the study's limitations), leading to our conclusion.

## II. METHODOLOGY

### A. Participants

The study protocol received approval from the SingHealth Centralised Institutional Review Board and all participants gave consent prior to joining the study. Participants comprised of 14 healthy adults with no known history of dysphagia, and 18 adult patients with dysphagia. All patients were recruited during their admission at Sengkang General Hospital (SKH). All dysphagic participants underwent a bedside swallow assessment followed by a Videofluoroscopic Swallowing Study (VFSS). Participants with a history of major head and/or neck surgery, presence of a tracheostomy tube, difficulty in sitting upright, displaying food refusal behaviours and have difficulty in following 1-step instructions consistently were excluded from the study.

### B. Signal Collection Protocol

Signal collection involved the use of the AeviceMD, a wearable stethoscope patented by Aevice Health. This device, a Singapore Health Science Authority Class B and United States Food and Drug Administration Class 2 Medical Device cleared for marketing, has previously demonstrated effectiveness in recording and analyzing lung sounds for early detection of chronic respiratory disease exacerbation [12]. For this study, the device's software was modified to record swallowing sounds. The wearable stethoscope was secured at a midline location superior to the thyroid notch with the silicone-based DuPont Liveo Soft Skin adhesive to seal the sensor. This positioning was established to avoid obstructing the radiographic view during VFSS while capturing swallowing events from a consistent location. The signal was unfiltered due to the lack of established upper limits for the bandwidth of swallowing sounds and was sampled in standard .WAV format (at a sampling rate of 4 kHz). Data collection commenced before the VFSS and continued throughout the procedure, synchronized with video-fluoroscopy imaging.

During the VFSS, patients were administered Level 0 thin fluids, Level 2 mildly thick fluids [13] and porridge (three teaspoons of each fluid consistency) by a Therapy Assistant as part of the swallowing evaluation. Swallows were performed with the participant in a neutral head position to optimize image capture on VFSS. Swallows involving compensatory strategies such as 3 second prep and effortful swallows were excluded from the analysis. A total of 152 normal swallows and 110 dysphagic swallows were collected.

Two experienced speech therapists with established inter-rater reliability reviewed the VFSS data to annotate swallow segments from the raw audio samples, ensuring synchronization with VFSS. These annotated swallow audio recordings, collected using the cervical auscultation protocol developed were then processed and analysed.

### C. Signal Processing, Analysis and Interpretation

The audio samples recorded during cervical auscultation underwent several processing steps to facilitate the development of the audio classification algorithm, including:

- Synchronization: audio recordings were synced with VFSS to identify swallowing events.
- Audio segmentation: Swallowing audio signals were segmented creating an audio database.
- Feature extraction: From the entire segments of swallowing audio signals, 25 audio features were extracted. These include 13 Mel Frequency Cepstral Coefficients (MFCC 1 to MFCC 13), other descriptors such as spectral centroid, spectral entropy, spectral flatness, spectral flux, spectral kurtosis, spectral roll off, spectral skewness, spectral spread, harmonic ratio, zero-cross rate, short term energy [14-15].
- Classification Modelling: Building on the insights from the previous analyses, we sought to assess the feasibility of using machine learning to detect dysphagia from swallowing sounds. The data (262 swallow sounds) was divided into training sets (60% or 157 swallow sounds) and testing sets (40% or 105 swallow sounds). We utilized the Random Forest Classifier [16], a well-established machine learning model known for its robustness, its ability to handle high-dimensional data, its efficacy in dealing with small datasets and its competence in handling imbalanced data. To address potential biases associated with a restricted sample size and enhance the robustness of our evaluations, we conducted 11 iterations of randomly dividing the data into separate training and testing sets, and aggregated the results across all iterations, thus ensuring the model's generalizability and reduces the risk of possible overfitting arising from any single data partition. We evaluated the performance of the trained model on independent testing sets using metrics such as accuracy, sensitivity, specificity, and F1 score.

## III. ANALYSES AND RESULTS

### A. Statistical Analysis of Acoustic Features

To assess whether there were significant differences in the acoustic features of swallowing sounds among all participants (including both healthy individuals and patients) across the different fluid and food consistencies, and underlying swallowing conditions, we employed a one-way ANOVA (non-parametric) - Kruskal Wallis test. This analysis aimed to establish a baseline understanding of the acoustic differences associated with the various consistencies.

No significant differences ($p>0.05$) were observed for all the acoustic features among different consistencies within each group (i.e., within healthy swallow group and within pathological swallow group). Figure 1 illustrates an example rain cloud plot of spectral crest distributions for various consistencies across healthy and pathological swallow groups. The distributions significantly overlap, suggesting similarity of this feature across the different consistencies swallowed.

When analyzing normal against pathological swallow sound signals, spectral crest, zero-crossing rate, MFCC5 and MFCC6 showed statistically significant differences ($p<0.05$) between the normal and pathological swallows (regardless of fluid consistency). Comparing healthy and pathological swallows, the rain cloud plot in Figure 2 shows minimal overlap in spectral crest and MFCC5 distributions (as examples), indicating differences in feature means.

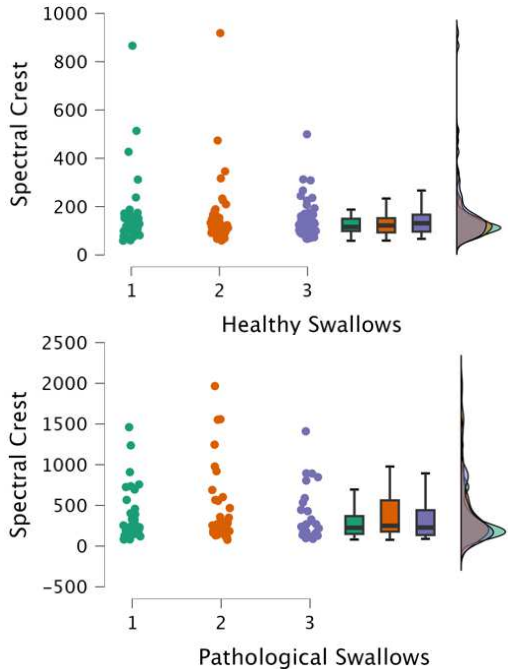

*Fig 1 Distribution of spectral crest in swallow sounds for various fluid consistencies swallowed: 1 – Level 2 mildly thick fluids, 2 – Porridge and 3 – Level 0 thin fluids, across healthy (top) and pathological (bottom) groups.*

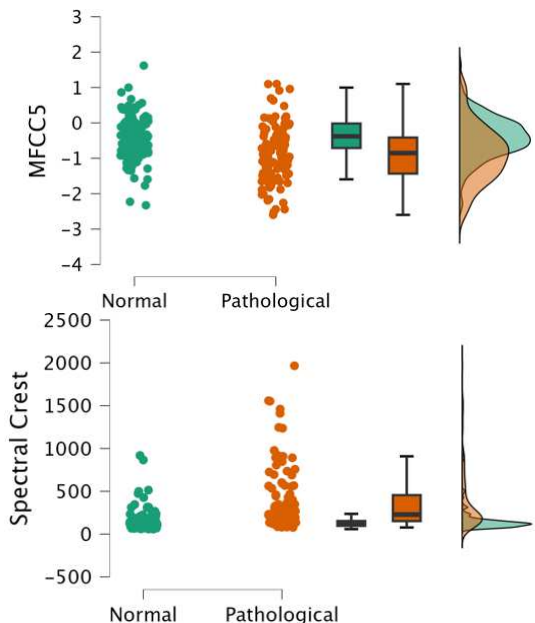

*Fig 2 Feature distribution comparison of healthy and pathological groups for MFCC5 (top) and Spectral crest (bottom).*

### B. Analysis 2: Feature visualisation

To gain insight into the separability of normal and dysphagic swallows in the acoustic feature space, 25 extracted acoustic features were transformed using principal component analysis (PCA) [17] and t-distributed stochastic neighbour embedding (t-SNE) [18]. These techniques were employed to reduce the high-dimensional feature space into lower dimensions and visualize the data. The transformation (Figure 3) revealed some separation between normal and dysphagic swallows in the lower dimensional space, indicating that the acoustic features are possibly carrying information that could potentially be exploited for classification purposes.

**TABLE I.** DYSPHAGIC CLASSIFICATION PERFORMANCE OF RANDOM FOREST CLASSIFIER (Mean ± Standard Deviation)

| Attribute | Dysphagic Swallows |
|---|---|
| True Positives | 33 ± 4 |
| False Positives | 7 ± 4 |
| False Negatives | 12 ± 4 |
| True Negatives | 56 ± 4 |
| Precision (%) | 84 ± 7 % |
| Sensitivity (%) | 74 ± 8 % |
| Specificity (%) | 89 ± 6 % |
| Accuracy (%) | 83 ± 3 % |
| F1 Score (%) | 78 ± 5 % |

### C. Analysis 3: Machine learning classification

The performance of Random Forest classification model aggregated over 11 iterations is presented in Table 1. The system demonstrated fair sensitivity (mean ± SD: 74% ± 8%) and specificity (89% ± 6%) for dysphagic swallows. Sensitivity measures the proportion of actual dysphagia swallows that are correctly identified by the classifier. A high sensitivity means that the classifier is effective at detecting dysphagia accurately (true positives), while specificity refers to the proportion of actual normal swallows that are correctly identified as not having dysphagia. A high specificity indicates that the classifier is good at identifying true negatives and not misclassifying them as dysphagic swallows. This is further reflected in the corresponding higher True Positives than False Negatives values, as well as higher True Negatives than False Negatives values, indicating the potential of the system in identifying pathological swallows.

Accuracy measures the overall ability of the classifier to correctly classify both dysphagia and normal swallows by considering both true positives and true negatives out of all classifications made. Here, the model attained an overall accuracy of 83% ± 3%. The F1 score of 78 ± 5 % indicates that the model is likely to have both fair precision (the ability to correctly identify those with dysphagia) and fair recall (the ability to identify most actual cases of dysphagia). However, given the serious nature of swallowing disorders, it will be desirable to reduce the variance and aim for an even higher F1 score to ensure consistent and reliable detection.

### D. Analysis 4: Feature Importance in classification

The importance of features for differentiating normal and dysphagic swallowing sounds made used by the random forest classifier was estimated. Figure 4 shows the importance of the involved features across the 11 distinct models.

Spectral crest, zero-cross rate, and MFCC5 were the three most important features, with median importance of 15.62%, 7.57% and 6.24%, respectively. Other features with relatively high median importance include MFCC6 (5.36%), Spectral Entropy (4.78%), and MFCC2 (4.64%). These features likely capture important information about the spectral and temporal characteristics of swallowing sounds relevant for differentiating between normal and dysphagic swallows. More importantly, these features exhibited statistically significant differences (p<0.05) between the healthy and pathological swallows (see Section III A). This further underscores their critical role in discriminating between normal and dysphagic swallowing sounds.

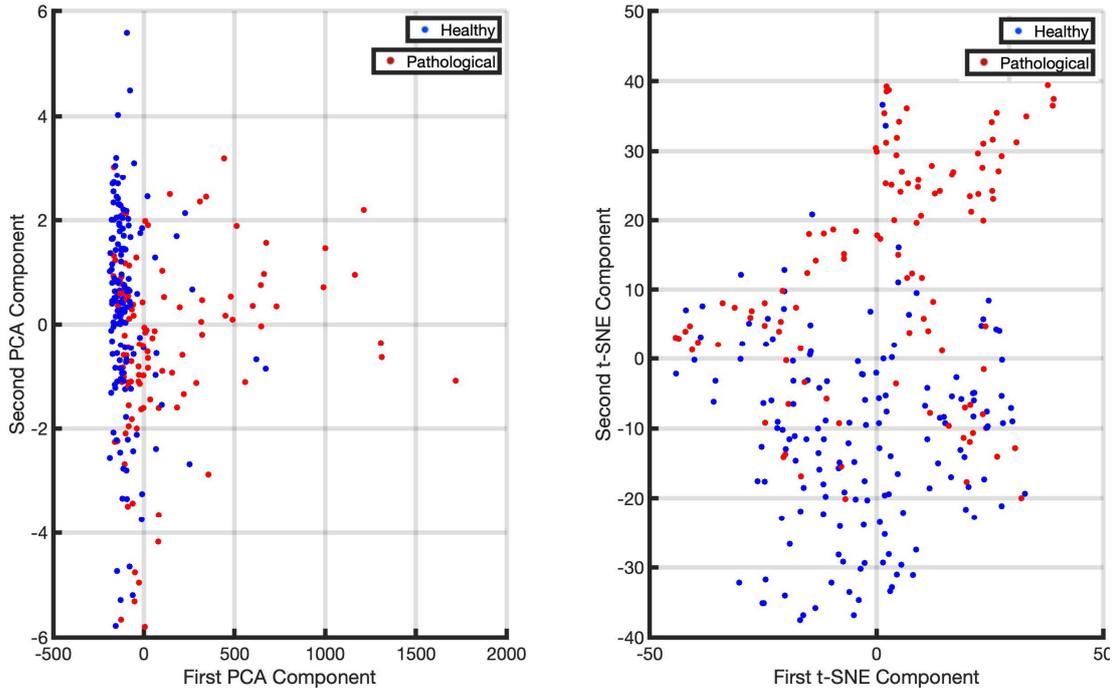

*Fig 3 Feature space distribution with PCA and t-SNE: Focusing on first two transformed components.*

## IV. Discussion

In this feasibility study, we show that acoustic features extracted from swallowing sounds can offer discriminative markers for detecting dysphagia, allowing for non-invasive, fast and relatively straightforward deployment in-the-field.

Acoustic features were seen to be associated consistently for audio signals from normative and dysphagic swallowing patterns, allowing for dimension reduction for data visualization using PCA and t-SNE approaches to indicate how demarcated the respective parameter space between normal and pathological swallows are. This provides the confidence to proceed with machine learning using the Random Forest Classifier, despite the challenge of noisy, dirty audio signals collected in-situ from the hospital setting, yielding an overall accuracy of 83% ± 3%, indicating the effectiveness of the acoustic approach, despite a fairly modest dataset (14 healthy participants contributing 152 swallows, 18 dysphagia patients providing 110 swallows).

Crucially, the nature of what was swallowed (different consistencies) did not affect the performance of the classification – in practical terms, this suggests acoustic assessment of dysphagia may not particularly favor any fluid consistency, potentially simplifying clinical deployment.

While the model shows promise, the variability in sensitivity and specificity suggests the current algorithm may not be consistently reliable for clinical use yet. In clinical practice, high sensitivity is crucial for screening tools to ensure that individuals with dysphagia are not missed. However, the lower end of this range (66%) might raise concerns for clinicians, as it suggests a significant proportion of dysphagia cases could be missed. Further, lower specificity might lead to overdiagnosis and inappropriate referrals, where normal swallows are incorrectly flagged as abnormal.

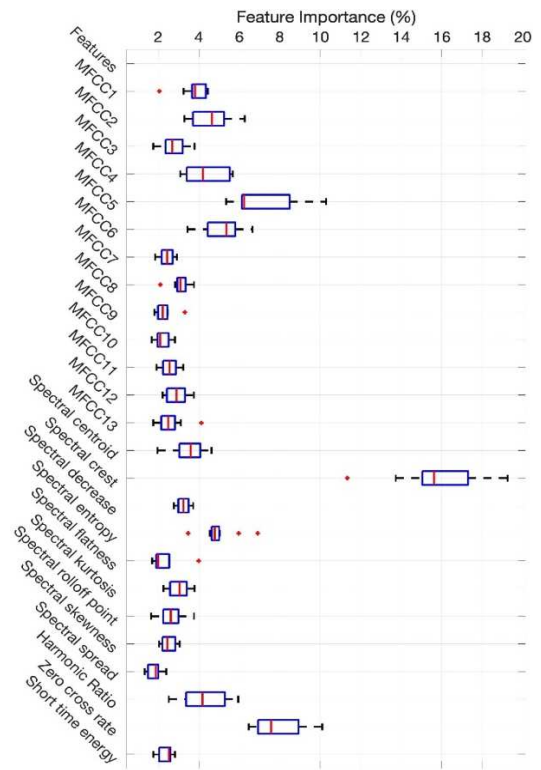

*Fig 4 The distribution of feature importance for acoustic features aggregated across 11 distinct models.*

It is also important to acknowledge our study's limitations:

- Modest sample size. The 14 healthy and 18 dysphagic participants may compromise the generalizability of the

results and the efficacy of the algorithm across a wider population may not be as representative.
- Sampling rate limitation. Hamlet et al [19] reported key spectral elements of swallowing sounds on fluids below 3 kHz, with higher frequencies overlapping with noise. Our device's 4 kHz sampling rate, capable of capturing frequencies up to 2 kHz, may potentially miss acoustic cues between 2 kHz and 3 kHz which may improve differentiating the types of swallows.
- Inter-subject and intra-subject variation. Variability in swallowing sounds inherently arise across individuals and within the same individual over time. Multiple swallows collected from each participant over time to better account for these variabilities will be beneficial.
- Machine learning optimization. This study was limited to conventional machine learning, which relied on judicious selection of appropriate signal features. In contrast, deep learning algorithms may offer improved performance by autonomously identifying inherent data features.

Accordingly, future work should consider the following:

- Analysis of swallowing sounds from a larger sample size with machine learning optimization. This would help to ascertain the sensitivity and specificity of this device being used as an assessment and screening tool for use in the community or in different healthcare settings where speech therapy services are not easily accessible for remote monitoring of dysphagia.
- Further classification of dysphagic patients into two categories for analysis - patients with no penetration and/or aspiration events vs patients with penetration and/or aspiration events. The analysis will include observing for any changes in acoustic signals that correspond to these clinical events. Knowing this correlation will allow for identification of dysphagic patients who are at high risk of silent penetration and/or aspiration so that a referral to a Speech Therapist can be made for timely assessment and management of their swallowing.
- Future research should consider the application of deep learning algorithms, which have the potential to automatically extract the most relevant features from the data. This could enhance the algorithm's ability to discern complex patterns yielding superior dysphagia screening performance.
- Conducting a comparative study between traditional machine learning models and deep learning models could provide valuable insights into the advantages and limitations of each approach within the context of swallowing sound analysis for screening dysphagia.

## V. Conclusion

Acoustic features offer fair differentiation between normal and pathological swallows for CA signals from a wearable stethoscope. Different fluid consistencies in both normal and pathological swallows yielded no significant differences in acoustic features. Despite the modest dataset, machine learning demonstrated promise in discriminating pathological and normal swallows. Such an approach may be useful as a screening tool in the community and a supplement to improving accuracy of bedside swallowing assessments, which can ultimately enhance patient care and outcomes. Further studies to include a larger sample size and further classification of dysphagic swallows may be considered.